**Quantifying human mobility behavior changes in response to non-pharmaceutical interventions during the COVID-19 outbreak in the United States.**

**Yixuan Pan, M.S., Aref Darzi, M.S., Aliakbar Kabiri, B.S., Guangchen Zhao, B.S., Weiyu Luo, M.S., Chenfeng Xiong, Ph.D., Lei Zhang, Ph.D.**

Department of Civil and Environmental Engineering
University of Maryland
1173 Glenn Martin Hall, College Park, MD 20742

Corresponding Author:

**Lei Zhang, Ph.D.**
Herbert Rabin Distinguished Professor
Director, Maryland Transportation Institute
Department of Civil and Environmental Engineering
University of Maryland
1173 Glenn Martin Hall, College Park, MD 20742
Email: lei@umd.edu
Phone: +1 (301) 405-2881

**Abstract**

Ever since the first case of the novel coronavirus disease (COVID-19) was confirmed in Wuhan, China, social distancing has been promoted worldwide, including the United States. It is one of the major community mitigation strategies, also known as non-pharmaceutical interventions. However, our understanding is remaining limited in how people practice social distancing. In this study, we construct a Social Distancing Index (SDI) to evaluate people's mobility pattern changes along with the spread of COVID-19. We utilize an integrated dataset of mobile device location data for the contiguous United States plus Alaska and Hawaii over a 100-day period from January 1, 2020 to April 9, 2020. The major findings are: 1) the declaration of the national emergency concerning the COVID-19 outbreak greatly encouraged social distancing and the mandatory stay-at-home orders in most states further strengthened the practice; 2) the states with more confirmed cases have taken more active and timely responses in practicing social distancing; 3) people in the states with fewer confirmed cases did not pay much attention to maintaining social distancing and some states, e.g., Wyoming, North Dakota, and Montana, already began to practice less social distancing despite the high increasing speed of confirmed cases; 4) some counties with the highest infection rates are not performing much social distancing, e.g., Randolph County and Dougherty County in Georgia, and some counties began to practice less social distancing right after the increasing speed of confirmed cases went down, e.g., in Blaine County, Idaho, which may be dangerous as well.

## Introduction

Since the World Health Organization (WHO) announced COVID-19 a pandemic, the infectious illness without effective vaccinations has threatened many countries and regions. As one of the major non-pharmaceutical interventions, social distancing, or physical distancing, is considered as an effective way to reduce COVID-19 infections. In the United States, government agencies have taken actions step by step to promote social distancing and mitigate the spread of COVID-19, such as educating the public the importance of social distancing, closing non-essential businesses, and issuing mandatory stay-at-home orders. The questions immediately come up. How do people react to the government actions and perform social distancing? How would these control measures influence the modeling of transmission dynamics? When could we lift such restrictions? We therefore propose a Social Distancing Index (SDI) based on mobile device location data to unveil people's mobility patterns in reaction to COVID-19 and social distancing policies. The objective of this study is to provide more insights into people's movements that could help policymaking in public health and accommodate epidemic modeling improvements.

People's actual behaviors in response to the interventions play an important role in modeling transmission dynamics. The existing studies regarding impact assessment of control measures mainly estimated related modeling parameters by Markov Chain Monte Carlo (MCMC)[1], utilized the simulation models to estimate contact network based on a synthetic population[2], estimated the contact patterns using survey data, modeling, and simulation[3–4], and collected people's behavior reactions through dedicated surveys[5]. We found that there lack timely contributions from real-world observations. Meanwhile, studies that evaluated the mobility changes during the pandemic from real-time and real-world observations mainly focuses on a single indicator: the distances traveled. The topics include the development of a social distancing scoreboard at the nation, state, and county level[6], the direct impact of stay-at-home mandates[7], and the mobility patterns by income distribution[8], etc[9–11]. A single metric, such as distances traveled, is not sufficient to capture the mobility changes and to portray people's performances in social distancing. Considering the various measurements of people's mobility patterns, such as number of trips made per person, and origin and destination matrix that displays the trips made between regions, an inclusive index is needed to simplify the information regarding different dimensions of human movements. An index also makes it easier for the stakeholders to communicate with each other[12], especially during this global challenge in alleviating the COVID-19 pandemic.

In this study, we incorporate five basic metrics to comprehensively evaluate people's behaviors in social distancing, e.g., number of personal trips (work and non-work) made daily and percentage of out-of-county trips. These metrics are generated from mobile device location data by data fusion and analytics. Mobile device location data is an emerging data source that provides insights into real-time human mobility patterns with its large sample size and continuous observations. Researchers have utilized such data to understand individual human mobility patterns[13], to understand the spreading patterns of mobile phone viruses[14], to explore social ties and link prediction[15], and to evaluate the impact of human mobility on epidemics[16–18]. In response to the COVID-19 pandemic, researchers have also discussed how mobile device data could help policymakers control infection, optimize policymaking, and evaluate the effectiveness of released policies[19,20] without overlooking the privacy issues about digital data[21]. We hereby introduce mobile device location data as an appropriate and functional data source for measuring the great impact of COVID-19 and facilitating public health researches based on real-world observations.

## Methods

### Sources of data

For this study, the research team created a data panel by integrating multiple mobile device data sources representing person and vehicle movements to improve the quality of the data. The data providers collected the first-party data from anonymized users for privacy protection. Next, we went through the data cleaning process, in which the consistency, accuracy, completeness, and timeliness of all observations were checked, and all suspicious observations got removed. We then clustered the location points into activity locations to identify home and work locations at the census block group (CBG) level for privacy protection. After that, we applied previously developed algorithms[22] to produce trip level information from the cleaned data panel, including trip origin and destination, travel distance, departure time, and arrival time. If anonymized individuals in the sample did not make any trip longer than 1·61 km from home for a calendar day, we considered them as the stay-at-home population. A multi-level weighting procedure including device-level and trip-level weights was employed to expand the sample to the entire population and to ensure the representativeness of the population at the nation, state, and county level. The results of the computational algorithms have been validated based on several independent datasets, such as National Household Travel Survey (NHTS) and American Community Survey (ACS), and peer reviewed by an external expert panel[22]. Finally, the derived mobility metrics were integrated with COVID-19 case data[23] and population data[24], and published in the University of Maryland COVID-19 Impact Analysis Platform[25]. The platform aggregates mobile device location data from more than 100 million devices across the nation at a monthly basis. Additional details can be found in another paper by the authors[22].

Generated from the mobile device location data from January 1 to April 9, 2020, the five basic mobility metrics are defined and summarized in Table 1. The basic metrics are selected to cover the frequency, spatial range, and semantics of people's daily travel.

**Table 1. Definition and Descriptive Statistics (State-level) for the Basic Metrics**

| Index | Metric | Description | Min | Max | Mean | Median |
|---|---|---|---|---|---|---|
| 1 | Percentage of residents staying home | Percentage of residents that make no trips more than 1·61 km away from home. | 13·0 | 57·0 | 23·7 SD: 7·1 | 21·00 |
| 2 | Daily work trips per person | Average number of work trips made per person. A work trip is a trip going to or from one's imputed work location. | 0·10 | 1·80 | 0·48 SD: 0·21 | 0·50 |
| 3 | Daily non-work trips per person | Average number of non-work trips made per person. | 1·60 | 3·90 | 2·76 SD: 0·36 | 2·80 |
| 4 | Distances traveled per person | Distances in kilometers traveled per person on all travel modes. | 17·2 | 104·3 | 58·2 SD: 13·6 | 59·4 |
| 5 | Out-of-county trips (in thousands) | Number of all trips that travels from and to the outside of the county. | 8 | 35917 | 6903 SD: 6961 | 4425 |

**Social Distancing Index**

In order to properly design the structure of the Social Distancing Index (SDI), we have reviewed the existing indices from various fields. Based on our findings, there are mainly two types of indices: category-based indices and score-based ones. The category-based indices explain the proposed objective by categories. For instance, Pandemic Severity Index (PSI) classifies the case fatality ratio (CFR) of a disease into five categories (from one to five)[26], and Modified Mercalli Intensity Scale evaluates the severity of an earthquake by categorizing it into twelve bins from I to XII[27]. On the other hand, score-based indices usually define a score from zero to one hundred to differentiate objectives and rank them in order. For example, US News State Ranking creates a score that covers eight topics on people's needs in each state and assigns different weights to those topics based on the survey data[28]. Bloomberg Global Health Index is another score-based index that ranks countries in terms of healthiness by giving them a rate between zero and one hundred[29].

It can be summarized that the category-based indices are usually built upon a single variable, and the score-based ones are more capable of integrating multiple metrics to be more informative. Therefore, we design SDI as a score-based one, which gives a 0-100 score to each geographical area, e.g., a state or a county, and measures to what extent the residents in the area and visitors to the area practice social distancing. Zero indicates no social distancing and one hundred indicates perfect social distancing compared with the benchmark days before the COVID-19 outbreak. The benchmark values for the basic metrics are computed using data from the weekdays (Monday to Friday) during the first two weeks of February. Thereafter, the changes in people's mobility patterns are captured by percentage reduction of the corresponding metrics in Table 1 (noted as $X_2, \ldots, X_5$) as input. And the absolute changes in the percentage of residents staying home (noted as $X_1$) also serve as input. The percentage reductions are absolute values between 0–100%. Any increase will be standardized as 0% in the calculation.

By jointly considering the travel behaviors of the region residents and visitors, the equation for computing SDI is given as follows.

$$SDI = [(\beta_1 X_1 + 0 \cdot 01 \times (100 - X_1) \times (\beta_2 X_2 + \beta_3 X_3 + \beta_4 X_4)] \times (1 - \beta_5) + \beta_5 X_5$$

Where $\beta_1 = 1$ and $\beta_2 + \beta_3 + \beta_4 = 1$.

The first part of the equation focuses on resident trips and the second part on out-of-county trips. $\beta_5$ is thus the weight assigned to behavior changes regarding out-of-county trips. For the resident trips, we use percentage of residents staying home to account for residents that do not make trips longer than $1 \cdot 61$ km from home, so the weight is simply one ($\beta_1 = 1$). For the people not staying home (travelers), percentage of which is $100 - X_1$, we use a weighted sum of percentage reductions in the number of work and non-work trips made daily, and the average distances traveled per person. When they make more work and non-work trips, and travel longer distances, they are considered to practice less social distancing. The weights for each variable should sum up to one ($\beta_2 + \beta_3 + \beta_4 = 1$) so that the resident travelers are comparable to residents staying at home.

To assign appropriate weights to each variable, we have consulted both actual observations and conceptual guidelines. Firstly, we observe that the relative ratio between resident trips and out-of-county trips nationwide is about four to one. Hence, we assign a weight of $0 \cdot 2$ to $\beta_5$. Secondly, it is widely observed that people have significantly reduced travel distances so the index should avoid the large percentage reduction in distances traveled to overwhelm the reductions in number of trips. Meanwhile, the reductions in number of trips made are more informative regarding people's reactions to the stay-at-home mandates. We thus consider the reduction in number of trips twice as important as that in distances traveled, and assign a weight of $0 \cdot 3$ to $\beta_4$. Moreover, as suggested by government agencies, people are highly encouraged to reduce non-essential trips. The index should be designed to favor the reduction in non-essential trips, which is estimated twice as important as the reduction in essential trips. The work trips are intuitively considered as essential trips and the non-work trips could include both. Based on the 2017 NHTS Travel Profile[30], the relative ratio between essential and non-essential non-work trips is approximately 1:2. Therefore, the relative ratio between the percentage reduction of work and non-work trips is $1:1 \cdot 67$. According to the constraint $\beta_2 + \beta_3 + \beta_4 = 1$, we further assign $0 \cdot 25$ to $\beta_2$ and $0 \cdot 45$ to $\beta_3$. In this study, SDI is eventually computed as follows.

$$SDI = [(X_1 + 0 \cdot 01 \times (100 - X_1) \times (0 \cdot 25 X_2 + 0 \cdot 45 X_3 + 0 \cdot 3 X_4)] \times 0 \cdot 8 + 0 \cdot 2 X_6$$

## Results

The proposed SDI is sensitive to people's behavior changes and is capable of reflecting the mobility changes accordingly. We examine the effectiveness and reasonableness of the proposed SDI by reviewing its temporal change for the entire nation (Figure 1). The SDI changes clearly indicate that people stay home more and travel less on weekends, especially on Sundays. During the study period (100 days from January 1 to April 9, 2020), people began to practice significantly more social distancing nationwide after President Trump declared the national emergency concerning the COVID-19 outbreak. The national emergency declaration immediately triggered people's responses on the following weekdays starting from March 16, and on the weekends of the following weeks, e.g., March 22, March 29, and April 5. In addition, the range of index became wider after March 16, indicating that people from different states were taking distinct responses to the national emergency announcement, which will be discussed in the following sections. After the week of March 23, we observe a general plateau in terms of the mean value and range of SDI. However, after April 6, there is a tendency that some states started to practice less social distancing. The possible reasons are twofold. First, people become less attentive to the outbreak as the outbreak persists. Moreover, the great economic impact is threatening the entire nation so that some people cannot afford to maintain social distance.

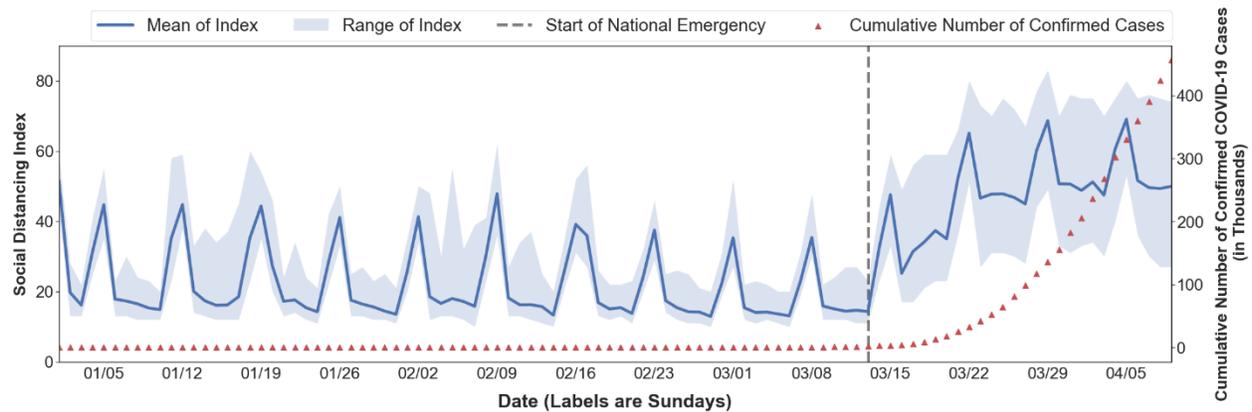

**Figure 1. Temporal changes of Social Distancing Index nationwide**

The mandatory stay-at-home orders issued by most states also triggered a second wave of strengthening social distancing, following the great impact of the declaration of the national emergency. However, in the week of April 6, the SDI scores of some states kept decreasing, such as Idaho (-9), Wyoming (-9), Montana (-8), and Alabama (-6). The SDI is computed for all states for seven consecutive weeks from February 20, 2020 to April 9, 2020 in Figure 2. The states are sorted in descending order by their SDI scores on the last day (April 9). The top five regions that are performing more social distancing are the District of Columbia, New York, New Jersey, Massachusetts, and Hawaii. Except for D.C., the other four states also issued the mandatory stay-at-home orders early. Meanwhile, the states practicing less social distancing are Wyoming, Arkansas, Idaho, Montana, Oklahoma, and South Dakota. Since most of them have not issued the stay-at-home mandates, people's performances in social distancing are almost consistent after the start of the national emergency. In the east and west coast, people intend to practice more social distancing potentially because they were exposed to the infection risk for a longer period and are aware of the higher infection risk with higher population density. Especially, the SDI scores kept increasing for the states in the northeast region in the week of April 6: Vermont (+8), New Hampshire (+6), Maine (+5), Rhode Island (+5), Massachusetts (+5), Connecticut (+4), New York (+3), Delaware (+2), etc. On the other hand, we observe continuously and significantly descending SDI scores in the same week for the following states: Idaho (-9), Wyoming (-9), Montana (-8), Alabama (-6), Alaska (-5),

Georgia (-5), Oregon (-5), North Carolina (-5), Tennessee (-5), Arkansas (-5), Washington (-4), and Arizona (-4).

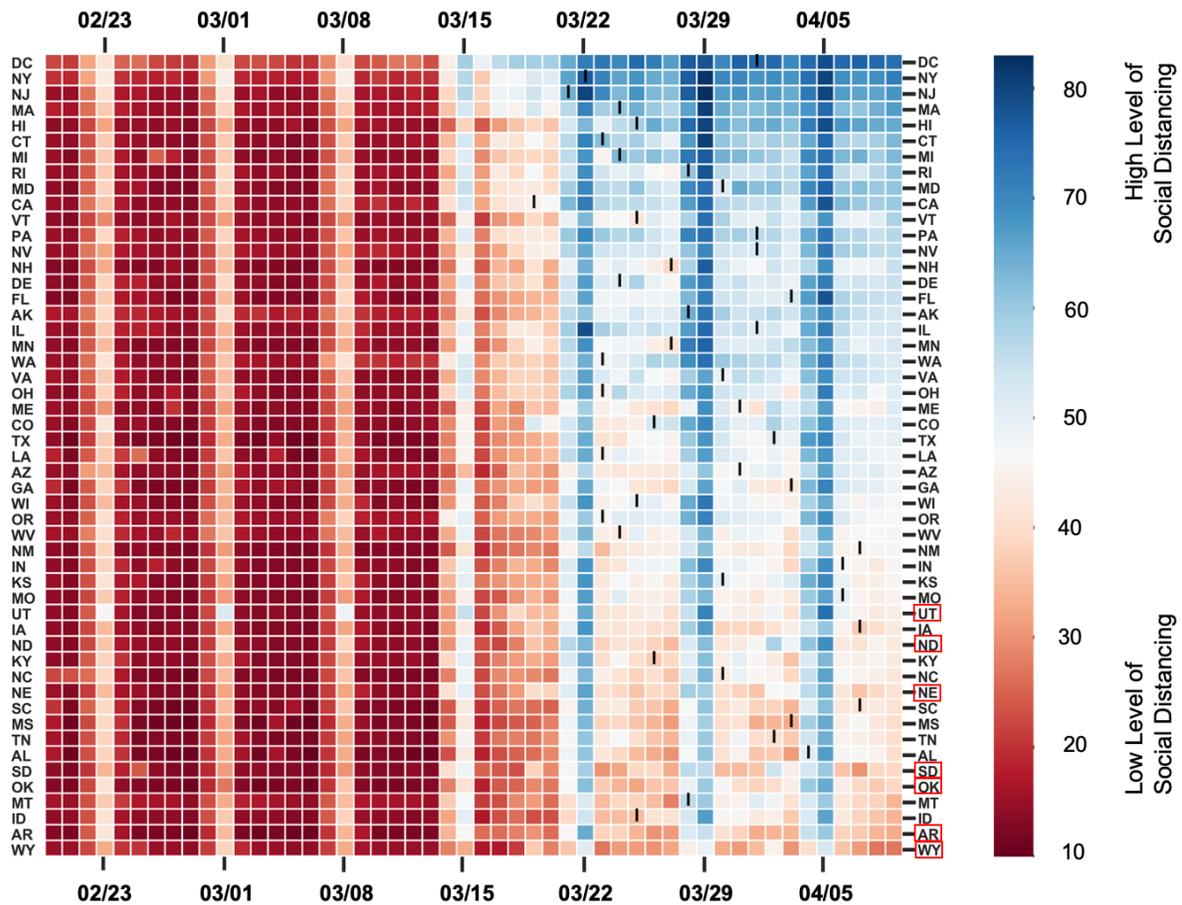

*The marker "I" indicates the start date of the mandatory stay-at-home order and the states without such orders are highlighted by red boxes. The dates shown are Sundays.*

**Figure 2. Social Distancing Index heatmap for all states.**

The states with more confirmed cases have taken more active and timely responses in social distancing and people are more responsive to the social distancing policies. However, the states with fewer infections have slower responses and people's behaviors did not change much in response to the mandatory stay-at-home orders. We have zoomed in on the top five and bottom five states regarding the cumulative number of confirmed cases on April 9 (Figure 3). After the stay-at-home orders were issued, all the top five states experienced a spike in SDI, but such orders have not encouraged much changes in the bottom five states, except Hawaii. It implies that the local severity of the COVID-19 outbreak plays a significant role in people's decision making. From the current observations, people in the top five states started conforming social distancing about one week before the local outbreak of COVID-19. Considering the time lag between infection and diagnosis, the benefits of social distancing may need more time to reveal. For the two states without stay-at-home mandates, the SDI increase is mainly prompted by the declaration of national emergency and the growth of confirmed cases. Although the bottom five states have fewer confirmed cases by the end of the study period (April 9), the risks of infection for these states are potentially higher than expected, especially in Wyoming, North Dakota, and Montana. People have not practiced more social distancing even after the confirmed cases started increasing at a fast pace. After April 6, people in these three states have practiced less social distancing, which is worth attention.

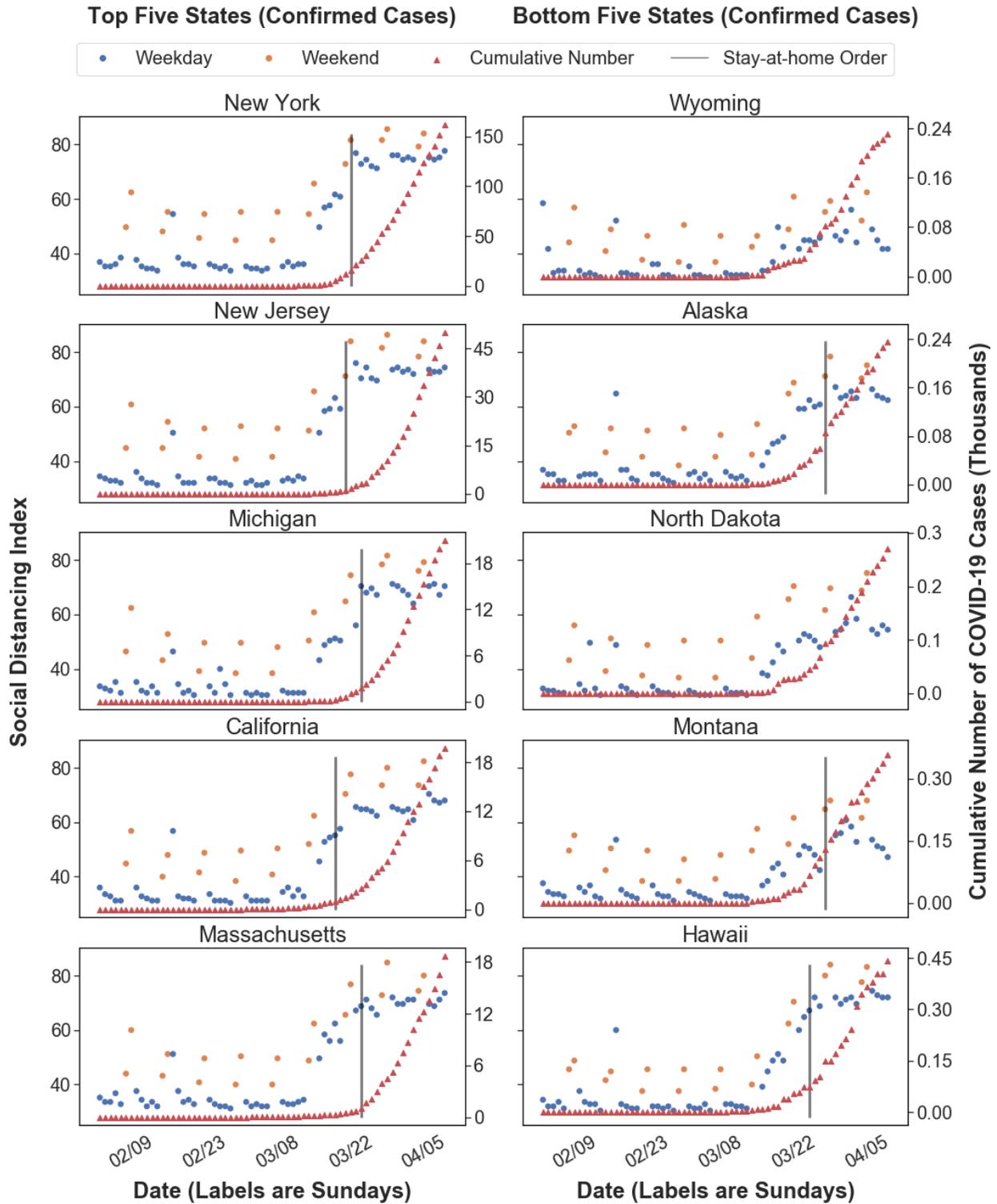

**Figure 3. Temporal changes of Social Distancing Index in the top five and bottom five states regarding the cumulative number of confirmed cases.**

SDI is also informative at the county level. Figure 4 demonstrates the temporal changes of SDI for the top ten counties regarding the cumulative number of confirmed cases per thousand people. The counties in New York are performing strict social distancing but Randolph County and Dougherty County in Georgia are not paying enough attention. Despite their high infection rates, their social distancing scores are relatively low, even after the stay-at-home mandate. The same concern exists in Blaine County, Idaho. People have practiced less social distancing right after the increasing speed of confirmed cases went down, which may result in a subsequent increase in transmission.

**Discussion**

During the crisis of the COVID-19 pandemic, data-driven tools that can provide insights into human behavior are of paramount importance. In this paper, we developed a Social Distancing Index (SDI) to capture the people's actual behaviors in social distancing by considering various aspects of human mobility patterns extracted from various sources of mobile device location data. Leveraging the merits of crowdsourcing, SDI is currently available at the nation, state, and county level[25] to help researchers quantify the influences of such community mitigation strategies. Monitoring the SDI patterns, both spatially and temporally, also enables policymakers to evaluate the effectiveness of related policies and to involve data-informed decision making for public health. In addition, SDI boosts the awareness of the general public and communities regarding the ongoing situation for where they are living. People can use insights from SDI to evaluate the potential risks in their neighborhoods.

Being exploratory research, this study could be further improved in several directions. Firstly, the basic mobility metrics could be generated considering the regional differences. Specifically, the current definition of staying at home population may introduce some bias due to different individual behaviors between residents in rural and urban areas. For example, many people living in rural regions still need to make long trips to shop for essential goods while people in urban areas have a higher chance to obtain essential items in a close neighborhood (within 1·61 km from home) and thus have a higher chance to be identified as staying at home. Secondly, adding more mobility metrics to the SDI could contribute to the comprehensiveness of the index. For instance, the trip purposes could be inferred by integrating mobile device location data and point of interest (POI) data. Identifying where people visit could allow us to distinguish between essential and non-essential trips, in addition to distinguishing between work and non-work trips. Thirdly, variables measuring the relationship between human movements and disease transmission could be extremely valuable. Although it may be difficult to retrieve details like contact tracing information from mobile device location data, the aggregate measurements can also be significant indicators, such as trips from and to the heavily infected areas that yield potential exposure and disease transmission in the study area, on top of out-of-county trips that are currently included. Moreover, an expert survey on improving the weight assignments to different variables in SDI may also contribute to a better construction of the index if time allows. Observing the mobility patterns and COVID-19 evolution for a longer period may also shed light on the assignment of weights.

Another future research direction is to integrate SDI with existing epidemiological frameworks, such as compartmental models. A variable of interest in these frameworks is to understand how the input variables evolve during the course of the outbreak. Certain policies such as mobility restrictions can significantly reduce input variables like reproduction factor of the disease. SDI can be employed in these models to enhance the input prediction in compartmental models.

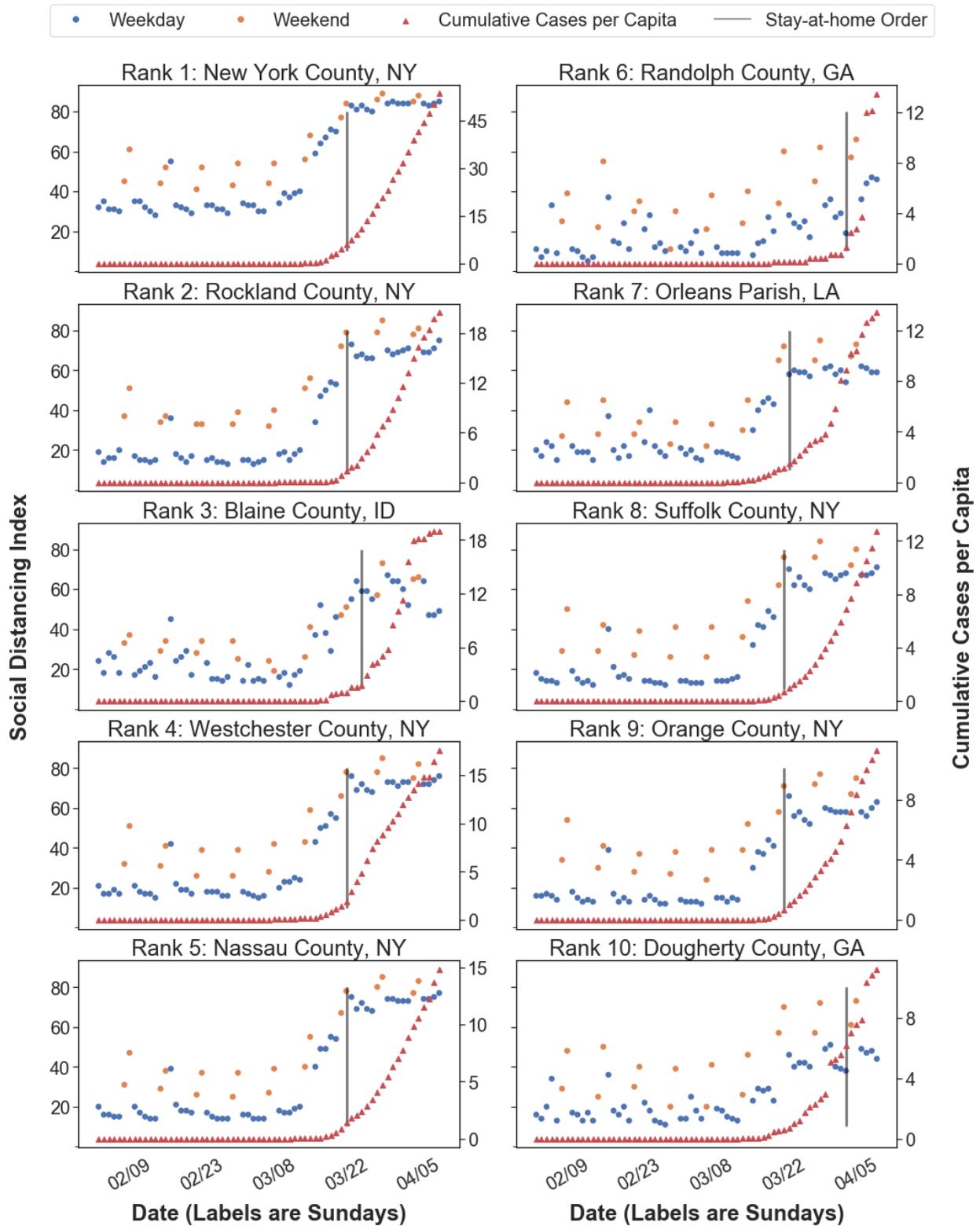

**Figure 4. Temporal changes of Social Distancing Index in the top ten counties regarding the cumulative number of confirmed cases per capita (per thousand people).**

## Contributors

LZ and YP designed the study. YP, AD, AK, GZ, WL and CX analyzed the data. YP, AD, AK, GZ, and WL interpreted the data. YP, AD, and AK wrote the first draft. All authors contributed to the final draft.

## Declaration of interests

All authors declare no competing interests.

## Acknowledgments

We would like to thank and acknowledge our partners and data sources in this effort: (1) partial financial support from the U.S. Department of Transportation's Bureau of Transportation Statistics, the National Science Foundation's RAPID Program, and Maryland Transportation Institute at the University of Maryland; (2) Amazon Web Service and its Senior Solutions Architect, Jianjun Xu, for providing cloud computing and technical support; (3) computational algorithms developed and validated in a previous USDOT Federal Highway Administration's Exploratory Advanced Research Program project; and (4) COVID-19 confirmed case data from the Johns Hopkins University Github repository and sociodemographic data from the U.S. Census Bureau.

22	Zhang L, Ghader S, Pack M, Xiong C, Darzi A, Yang M, Sun Q, Kabiri A, and Hu, S. An interactive COVID-19 mobility impact and social distancing analysis platform. Working Paper, Maryland Transportation Institute, University of Maryland, 2020. *medRxiv*.

23	Johns Hopkins University. COVID-19 Dashboard by the Center for Systems Science and Engineering (CSSE) at Johns Hopkins University. 2020. https://coronavirus.jhu.edu/map.html (accessed April 10, 2020).

24	U.S. Census Bureau. Annual Estimates of the Resident Population for Selected Age Groups by Sex for the United States, States, Counties and Puerto Rico Commonwealth and Municipios; using data.census.gov; published online June 2019. Accessed on March 30, 2020.

25	Maryland Transportation Institute. University of Maryland COVID-19 Impact Analysis Platform. 2020. https://data.covid.umd.edu (accessed on April 18, 2020).

26	Qualls N, Levitt A, Kanade N, et al. Community mitigation guidelines to prevent pandemic influenza—United States, 2017. *MMWR Recommendations and Reports* 2017; **66:** 1–34.

27	Wood HO, Neumann F. Modified Mercalli intensity scale of 1931. *Bulletin of the Seismological Society of America* 1931; **21:** 277–83.

28	U.S. News & World Report. Best states 2019: how they were ranked. Published online May 14, 2019. https://www.usnews.com/news/best-states/articles/methodology (accessed on April 18, 2020).

29	World Population Review. Healthiest Countries Population. Published online April 6, 2020. http://worldpopulationreview.com/countries/healthiest-countries/ (accessed on April 18, 2020).

30	Federal Highway Administration. 2017 National Household Travel Survey Travel Profile: United States. Washington, D.C.: U.S. Department of Transportation, 2019. https://nhts.ornl.gov/assets/2017_USTravelProfile.pdf (accessed on April 11, 2020).12